\documentstyle[12pt,color,epsfig,axodraw,subfigure]{article}
\def\baselinestretch{1.3}
\advance\textheight by \topskip
\newcommand{\comment}[1]{}

\def\beq{\begin{equation}}
\def\eeq{\end{equation}}
\def\beqn{\begin{eqnarray}}
\def\eeqn{\end{eqnarray}}

\begin{document}
\tolerance=100000
\thispagestyle{empty}
\setcounter{page}{0}
\topmargin -0.1in
\headsep 30pt
\footskip 40pt
\oddsidemargin 12pt
\evensidemargin -16pt
\textheight 8.5in
\textwidth 6.5in
\parindent 20pt
 
\def\baselinestretch{1.5}
\newcommand{\newc}{\newcommand}
\def\preprint{{preprint}}
\def\Ord{\lower .7ex\hbox{$\;\stackrel{\textstyle <}{\sim}\;$}}
\def\OOrd{\lower .7ex\hbox{$\;\stackrel{\textstyle >}{\sim}\;$}}
\def\cO#1{{\cal{O}}\left(#1\right)}
\newc{\order}{{\cal O}}
\def\lag             {{\cal L}}
\def\Lag             {{\cal L}}
\def\lum             {{\cal L}}
\def\R               {{\cal R}}
\def\Rsq             {{\cal R}^{\sq}}
\def\Rst             {{\cal R}^{\st}}
\def\Rsb             {{\cal R}^{\sb}}
\def\M               {{\cal M}}
\def\Oas             {{\cal O}(\alpha_{s})}
\def\Vcal            {{\cal V}}
\def\Wcal            {{\cal W}}
\newc{\be}{\begin{equation}}
\newc{\ee}{\end{equation}}
\newc{\br}{\begin{eqnarray}}
\newc{\er}{\end{eqnarray}}
\newc{\ba}{\begin{array}}
\newc{\ea}{\end{array}}
\newc{\bi}{\begin{itemize}}
\newc{\ei}{\end{itemize}}
\newc{\bn}{\begin{enumerate}}
\newc{\en}{\end{enumerate}}
\newc{\bc}{\begin{center}}
\newc{\ec}{\end{center}}
\newc{\ul}{\underline}
\newc{\ol}{\overline}
\newc{\ra}{\rightarrow}
\newc{\lra}{\longrightarrow}
\newc{\wt}{\widetilde}
\newc{\til}{\tilde}
\def\kr              {^{\dagger}}
\newc{\wh}{\widehat}
\newc{\ti}{\times}
\newc{\Dir}{\kern -6.4pt\Big{/}}
\newc{\Dirin}{\kern -10.4pt\Big{/}\kern 4.4pt}
\newc{\DDir}{\kern -10.6pt\Big{/}}
\newc{\DGir}{\kern -6.0pt\Big{/}}
\newc{\sig}{\sigma}
\newc{\sigmalstop}{\sig_{\lstoppair}}
\newc{\Sig}{\Sigma}  
\newc{\del}{\delta}
\newc{\Del}{\Delta}
\newc{\lam}{\lambda}
\newc{\Lam}{\Lambda}
\newc{\gam}{\gamma}
\newc{\Gam}{\Gamma}
\newc{\eps}{\epsilon}
\newc{\Eps}{\Epsilon}
\newc{\kap}{\kappa}
\newc{\Kap}{\Kappa}
\newc{\modulus}[1]{\left| #1 \right|}
\newc{\eq}[1]{(\ref{eq:#1})}
\newc{\eqs}[2]{(\ref{eq:#1},\ref{eq:#2})}
\newc{\etal}{{\it et al.}\ }
\newc{\ibid}{{\it ibid}.}
\newc{\ibidem}{{\it ibidem}.}
\newc{\eg}{{\it e.g.}\ }
\newc{\ie}{{\it i.e.}\ }
\def \viz{\emph{viz.}}
\def \etc{\emph{etc. }}
\newc{\nonum}{\nonumber}
\newc{\lab}[1]{\label{eq:#1}}
\newc{\dpr}[2]{({#1}\cdot{#2})}
\newc{\lt}{\stackrel{<}}
\newc{\gt}{\stackrel{>}}
\newc{\lsimeq}{\stackrel{<}{\sim}}
\newc{\gsimeq}{\stackrel{>}{\sim}}
\def\lsim{\buildrel{\scriptscriptstyle <}\over{\scriptscriptstyle\sim}}
\def\gsim{\buildrel{\scriptscriptstyle >}\over{\scriptscriptstyle\sim}}
\def\lapp{\mathrel{\rlap{\raise.5ex\hbox{$<$}}
                    {\lower.5ex\hbox{$\sim$}}}}
\def\gapp{\mathrel{\rlap{\raise.5ex\hbox{$>$}}
                    {\lower.5ex\hbox{$\sim$}}}}
\newc{\half}{\frac{1}{2}}
\newcommand {\nnc}        {{\overline{\mathrm N}_{95}}}
\newcommand {\dm}         {\Delta m}
\newcommand {\dM}         {\Delta M}
\def\bra{\langle}
\def\ket{\rangle}
\def\cO#1{{\cal{O}}\left(#1\right)}
\def \DM{{\Delta{m}}}
\newc{\bQ}{\ol{Q}}
\newc{\dota}{\dot{\alpha }}
\newc{\dotb}{\dot{\beta }}
\newc{\dotd}{\dot{\delta }}
\newc{\nindnt}{\noindent}

\newcommand{\medf}[2] {{\footnotesize{\frac{#1}{#2}} }}
\newcommand{\smaf}[2] {{\textstyle \frac{#1}{#2} }}
\def\onesq            {{\textstyle \frac{1}{\sqrt{2}} }}
\def\onehf            {{\textstyle \frac{1}{2} }}
\def\oneth            {{\textstyle \frac{1}{3} }}
\def\twoth            {{\textstyle \frac{2}{3} }}
\def\onefo            {{\textstyle \frac{1}{4} }}
\def\forth            {{\textstyle \frac{4}{3} }}

\newc{\matth}{\mathsurround=0pt}
\def\ML{\ifmmode{{\mathaccent"7E M}_L}
             \else{${\mathaccent"7E M}_L$}\fi}
\def\MR{\ifmmode{{\mathaccent"7E M}_R}
             \else{${\mathaccent"7E M}_R$}\fi}
\newcommand{\s}{\\ \vspace*{-3mm} }

\def \ud { {1 \over 2} }
\def \ut { {1 \over 3} }
\def \td { {3 \over 2} }
\newc{\mr}{\mathrm}
\def\dh {\partial }
\def \cs { cross-section }
\def \css { cross-sections }
\def \cm { centre of mass }
\def \cms { centre of mass energy }
\def \cc { coupling constant }
\def \ccs {coupling constants }
\def \gc {gauge coupling }
\def \gcc {gauge coupling constant }
\def \gccs {gauge coupling constants }
\def \yc {Yukawa coupling }
\def \ycc {Yukawa coupling constant }
\def \pp {{parameter }}
\def \pps {{parameters }} 
\def \ps {parameter space }
\def \pss {parameter spaces }
\def \vv {vice versa }

\newc{\siminf}{\mbox{$_{\sim}$ {\small {\hspace{-1.em}{$<$}}}    }}
\newc{\simsup}{\mbox{$_{\sim}$ {\small {\hspace{-1.em}{$>$}}}    }}


\newc {\Zboson}{{\mathrm Z}^{0}}
\newc{\thetaw}{\theta_W}
\newc{\mbot}{{m_b}}
\newc{\mtop}{{m_t}}
\newc{\sm}{${\cal {SM}}$}
\newc{\as}{\alpha_s}
\newc{\aem}{\alpha_{em}}
\def \PI{{\pi^{\pm}}}
\newc{\ppbar}{\mbox{$p\ol{p}$}}
\newc{\bbbar}{\mbox{$b\ol{b}$}}
\newc{\ccbar}{\mbox{$c\ol{c}$}}
\newc{\ttbar}{\mbox{$t\ol{t}$}}
\newc{\eebar}{\mbox{$e\ol{e}$}}
\newc{\zzero}{\mbox{$Z^0$}}
\def \gamz{\Gam_Z}
\newc{\wplus}{\mbox{$W^+$}}
\newc{\wminus}{\mbox{$W^-$}}
\newc{\ellp}{\ell^+}
\newc{\ellm}{\ell^-}
\newc{\elp}{\mbox{$e^+$}}
\newc{\elm}{\mbox{$e^-$}}
\newc{\elpm}{\mbox{$e^{\pm}$}}
\newc{\qbar}     {\mbox{$\ol{q}$}}
\def \ewgroup{SU(2)_L \otimes U(1)_Y}
\def \smgroup{SU(3)_C \otimes SU(2)_L \otimes U(1)_Y}
\def \smcolorem{SU(3)_C \otimes U(1)_{em}}

\def \SSM  {Supersymmetric Standard Model}
\def \poincare{Poincare$\acute{e}$}
\def \superspace{\emph{superspace}}
\def \sfs{\emph{superfields}}
\def \superpot{\emph{superpotential}}
\def \csf{\emph{chiral superfield}}
\def \csfs{\emph{chiral superfields}}
\def \vsf{\emph{vector superfield }}
\def \vsfs{\emph{vector superfields}}
\newc{\Ebar}{{\bar E}}
\newc{\Dbar}{{\bar D}}
\newc{\Ubar}{{\bar U}}
\newc{\susy}{{{SUSY}}}
\newc{\msusy}{{{M_{SUSY}}}}

\def\photino{\ifmmode{\mathaccent"7E \gam}\else{$\mathaccent"7E \gam$}\fi}
\def\taugluino{\ifmmode{\tau_{\mathaccent"7E g}}
             \else{$\tau_{\mathaccent"7E g}$}\fi}
\def\mphotino{\ifmmode{m_{\mathaccent"7E \gam}}
             \else{$m_{\mathaccent"7E \gam}$}\fi}
\newc{\gl}   {\mbox{$\wt{g}$}}
\newc{\mgl}  {\mbox{$m_{\gl}$}}
\def \charginopm{{\wt\chi}^{\pm}}
\def \mcharginopm{m_{\charginopm}}
\def \mchpmmin {\mcharginopm^{min}}
\def \chonep {{\wt\chi_1^+}}
\def \ch2p {{\wt\chi_2^+}}
\def \chonem {{\wt\chi_1^-}}
\def \ch2m {{\wt\chi_2^-}}
\def \chplus {{\wt\chi^+}}
\def \chminus {{\wt\chi^-}}
\def \chonip{{\wt\chi_i}^{+}}
\def \chonim{{\wt\chi_i}^{-}}
\def \chonipm{{\wt\chi_i}^{\pm}}
\def \chonjp{{\wt\chi_j}^{+}}
\def \chonjm{{\wt\chi_j}^{-}}
\def \chonjpm{{\wt\chi_j}^{\pm}}
\def \chonepm{{\wt\chi_1}^{\pm}}
\def \chonemp{{\wt\chi_1}^{\mp}}
\def \mchonepm{m_{\chonepm}}
\def \mchonemp{m_{\chonemp}}
\def \chtwopm{{\wt\chi_2}^{\pm}}
\def \mchtwopm{m_{\chtwopm}}
\newc{\dmchi}{\Delta m_{\wt\chi}}


\def \vlsp{\emph{VLSP}}
\def \lspi{\wt\chi_i^0}
\def \mlspi{m_{\lspi}}
\def \lspj{\wt\chi_j^0}
\def \mlspj{m_{\lspj}}
\def \lspone{\wt\chi_1^0}
\def \mlspone{m_{\lspone}}
\def \lsptwo{\wt\chi_2^0}
\def \mlsptwo{m_{\lsptwo}}
\def \lspthree{\wt\chi_3^0}
\def \mlspthree{m_{\lspthree}}
\def \lspfour{\wt\chi_4^0}
\def \mlspfour{m_{\lspfour}}


\newc{\sele}{\wt{\mathrm e}}
\newc{\sell}{\wt{\ell}}
\def \msell{m_{\sell}}
\def \slepone{\wt\ell_1}
\def \mslepone{m_{\slepone}}
\def \smuone{\wt\mu_1}
\def \msmuone{m_{\smuone}}
\def \stauone{\wt\tau_1}
\def \mstauone{m_{\stauone}}
\def \snu{\wt{\nu}}
\def \snutau{\wt{\nu}_{\tau}}
\def \msnu{m_{\snu}}
\def \msnumu{m_{\snu_{\mu}}}
\def \barsnu{\wt{\bar{\nu}}}
\def \barsnul{\barsnu_{\ell}}
\def \snul{\snu_{\ell}}
\def \mbarsnu{m_{\barsnu}}
\newc{\snue}     {\mbox{$ \wt{\nu_e}$}}
\newc{\smu}{\wt{\mu}}
\newc{\stau}{\wt{\tau}}
\newc {\nuL} {\wt{\nu}_L}
\newc {\nuR} {\wt{\nu}_R}
\newc {\snub} {\bar{\wt{\nu}}}
\newc {\eL} {\wt{e}_L}
\newc {\eR} {\wt{e}_R}
\def \slepl{\wt{l}_L}
\def \mslepl{m_{\slepl}}
\def \slepr{\wt{l}_R}
\def \mslepr{m_{\slepr}}
\def \stau{\wt\tau}
\def \mstau{m_{\stau}}
\def \slepton{\wt\ell}
\def \mslepton{m_{\slepton}}
\def \mlhiggs{m_{h^0}}

\def \xr{X_{r}}

\def \sfer{\wt{f}}
\def \msfer{m_{\sfer}}
\def \sq{\wt{q}}
\def \msq{m_{\sq}}
\def \msquleft{m_{\tilde{u_L}}}
\def \msqurht{m_{\tilde{u_R}}}
\def \sql{\wt{q}_L}
\def \msql{m_{\sql}}
\def \sqr{\wt{q}_R}
\def \msqr{m_{\sqr}}
\newc{\msqot}  {\mbox{$m_(\sq_{1,2} )$}}
\newc{\sqbar}    {\mbox{$\bar{\wt{q}}$}}
\newc{\ssb}      {\mbox{$\squark\ol{\squark}$}}
\newc {\qL} {\wt{q}_L}
\newc {\qR} {\wt{q}_R}
\newc {\uL} {\wt{u}_L}
\newc {\uR} {\wt{u}_R}
\def \ul{\wt{u}_L}
\def \mul{m_{\ul}}
\newc {\dL} {\wt{d}_L}
\newc {\dR} {\wt{d}_R}
\newc {\cL} {\wt{c}_L}
\newc {\cR} {\wt{c}_R}
\newc {\sL} {\wt{s}_L}
\newc {\sR} {\wt{s}_R}
\newc {\tL} {\wt{t}_L}
\newc {\tR} {\wt{t}_R}
\newc {\stb} {\ol{\wt{t}}_1}
\newc {\sbot} {\wt{b}_1}
\newc {\msbot} {m_{\sbot}}
\newc {\sbotb} {\ol{\wt{b}}_1}
\newc {\bL} {\wt{b}_L}
\newc {\bR} {\wt{b}_R}
\def \mul{m_{\wt{u}_L}}
\def \mur{m_{\wt{u}_R}}
\def \mdl{m_{\wt{d}_L}}
\def \mdr{m_{\wt{d}_R}}
\def \mcl{m_{\wt{c}_L}}
\def \charml{\wt{c}_L}
\def \mcr{m_{\wt{c}_R}}
\newc{\csquark}  {\mbox{$\wt{c}$}}
\newc{\csquarkl} {\mbox{$\wt{c}_L$}}
\newc{\mcsl}     {\mbox{$m(\csquarkl)$}}
\def \msl{m_{\wt{s}_L}}
\def \msr{m_{\wt{s}_R}}
\def \mbl{m_{\wt{b}_L}}
\def \mbr{m_{\wt{b}_R}}
\def \mtl{m_{\wt{t}_L}}
\def \mtr{m_{\wt{t}_R}}
\def \st{\wt{t}}
\def \mst{m_{\st}}
\newc {\stopl}         {\wt{\mathrm{t}}_{\mathrm L}}
\newc {\stopr}         {\wt{\mathrm{t}}_{\mathrm R}}
\newc {\stoppair}      {\wt{\mathrm{t}}_{1}
\bar{\wt{\mathrm{t}}}_{1}}
\def \lstop{\wt{t}_{1}}
\def \lstopbar{\lstop^*}
\def \hstop{\wt{t}_{2}}
\def \hstopbar{\hstop^*}
\def \mlstop{m_{\lstop}}
\def \mhstop{m_{\hstop}}
\def \lstoppair{\lstop\lstop^*}
\def \hstoppair{\hstop\hstop^*}
\newc{\tsquark}  {\mbox{$\wt{t}$}}
\newc{\ttb}      {\mbox{$\tsquark\ol{\tsquark}$}}
\newc{\ttbone}   {\mbox{$\tsquark_1\ol{\tsquark}_1$}}
\def \tsq {top squark }
\def \tsqs {top squarks }
\def \tsql {ligtest top squark }
\def \tsqh {heaviest top squark }
\newc{\mix}{\theta_{\wt t}}
\newc{\cost}{\cos{\theta_{\wt t}}}
\newc{\sint}{\sin{\theta_{\wt t}}}
\newc{\costloop}{\cos{\theta_{\wt t_{loop}}}}
\def \lsbot{\wt{b}_{1}}
\def \lsbotbar{\lsbot^*}
\def \hsbot{\wt{b}_{2}}
\def \hsbotbar{\hsbot^*}
\def \mlsbot{m_{\lsbot}}
\def \mhsbot{m_{\hsbot}}
\def \lsbotpair{\lsbot\lsbot^*}
\def \hsbotpair{\hsbot\hsbot^*}
\newc{\mixsbot}{\theta_{\wt b}}

\def \mhone{m_{h_1}}
\def \hup{{H_u}}
\def \hdn{{H_d}}
\newc{\tb}{\tan\beta}
\newc{\cb}{\cot\beta}
\newc{\vev}[1]{{\left\langle #1\right\rangle}}

\def \abot{A_{b}}
\def \atop{A_{t}}
\def \atau{A_{\tau}}
\newc{\mhalf}{m_{1/2}}
\newc{\mzero} {\mbox{$m_0$}}
\newc{\azero} {\mbox{$A_0$}}

\newc{\lb}{\lam}
\newc{\lbp}{\lam^{\prime}}
\newc{\lbpp}{\lam^{\prime\prime}}
\newc{\rpv}{{\not \!\! R_p}}
\newc{\rpvm}{{\not  R_p}}
\newc{\rp}{R_{p}}
\newc{\rpmssm}{{RPC MSSM}}
\newc{\rpvmssm}{{RPV MSSM}}


\newc{\sbyb}{S/$\sqrt B$}
\newc{\pelp}{\mbox{$e^+$}}
\newc{\pelm}{\mbox{$e^-$}}
\newc{\pelpm}{\mbox{$e^{\pm}$}}
\newc{\epem}{\mbox{$e^+e^-$}}
\newc{\lplm}{\mbox{$\ell^+\ell^-$}}
\def \branch{\emph{BR}}
\def \branche{\branch(\lstop\ra be^{+}\nu_e \lspone)\ti \branch(\lstop^{*}\ra \bar{b}q\bar{q^{\prime}}\lspone)}
\def \branchmu{\branch(\lstop\ra b\mu^{+}\nu_{\mu} \lspone)\ti \branch(\lstop^{*}\ra \bar{b}q\bar{q^{\prime}}\lspone)}
\def\Ecm{\ifmmode{E_{\mathrm{cm}}}\else{$E_{\mathrm{cm}}$}\fi}
\newc{\rts}{\sqrt{s}}
\newc{\rtshat}{\sqrt{\hat s}}
\newc{\gev}{\,GeV}
\newc{\mev}{~{\rm MeV}}
\newc{\tev}  {\mbox{$\;{\rm TeV}$}}
\newc{\gevc} {\mbox{$\;{\rm GeV}/c$}}
\newc{\gevcc}{\mbox{$\;{\rm GeV}/c^2$}}
\newc{\intlum}{\mbox{${ \int {\cal L} \; dt}$}}
\newc{\call}{{\cal L}}
\def \met  {\mbox{${E\!\!\!\!/_T}$}}
\def \cpv  {\mbox{${CP\!\!\!\!/}$}}
\newc{\ptmiss}{/ \hskip-7pt p_T}
\def \eslash{\not \! E}
\def \etslash{\not \! E_T }
\def \ptslash{\not \! p_T }
\newc{\PT}{\mbox{$p_T$}}
\newc{\ET}{\mbox{$E_T$}}
\newc{\dedx}{\mbox{${\rm d}E/{\rm d}x$}}
\newc{\ifb}{\mbox{${\rm fb}^{-1}$}}
\newc{\ipb}{\mbox{${\rm pb}^{-1}$}}
\newc{\pb}{~{\rm pb}}
\newc{\fb}{~{\rm fb}}
\newc{\ycut}{y_{\mathrm{cut}}}
\newc{\chis}{\mbox{$\chi^{2}$}}
\def \hadron{\emph{hadron}}
\def \nlc{\emph{NLC }}
\def \lhc{\emph{LHC }}
\def \cdf{\emph{CDF }}
\def\dzero{\emptyset}
\def \tevatron{\emph{Tevatron }}
\def \lep{\emph{LEP }}
\def \jets{\emph{jets }}
\def \jet(s){\emph{jet(s) }}

\def\Crs{stroke [] 0 setdash exch hpt sub exch vpt add hpt2 vpt2 neg V currentpoint stroke 
hpt2 neg 0 R hpt2 vpt2 V stroke}
\def\loopdk{\lstop \ra c \lspone}
\def\brloopdk{\branch(\loopdk)}
\def\fourdk{\lstop \ra b \lspone  f \bar f'}
\def\brfourdk{\branch(\fourdk)}
\def\fourdklep{\lstop \ra b \lspone  \ell \nu_{\ell}}
\def\fourdkhad{\lstop \ra b \lspone  q \bar q'}
\def\brfourdklep{\branch(\fourdklep)}
\def\brfourdkhad{\branch(\fourdkhad)}
\def\twodk{\lstop \ra b \chonep}
\def\brtwodk{\branch(\twodk)}
\def\threedkslep{\lstop \ra b \wt{\ell} \nu_{\ell}}
\def\brthreedkslep{\branch(\threedkslep)}
\def\threedksnu{\lstop \ra b \wt{\nu_{\ell}} \ell }
\def\brthreedksnu{\branch(\threedksnu) }
\def\threedklsp{\lstop \ra b W \lspone }
\def\brthreedklsp{\\branch(\threedklsp) }
\def\topdk{t \ra \lstop \lspone}
\def\rpvdk{\lstop \ra e^+ d}
\def\brrpvdk{\branch(\rpvdk)}
\def\fonec{f_{11c}} 
\newc{\mpl}{M_{\rm Pl}}
\newc{\mgut}{M_{GUT}}
\newc{\mw}{M_{W}}
\newc{\mweak}{M_{weak}}
\newc{\mz}{M_{Z}}

\newc{\OPALColl}   {OPAL Collaboration}
\newc{\ALEPHColl}  {ALEPH Collaboration}
\newc{\DELPHIColl} {DELPHI Collaboration}
\newc{\XLColl}     {L3 Collaboration}
\newc{\JADEColl}   {JADE Collaboration}
\newc{\CDFColl}    {CDF Collaboration}
\newc{\DXColl}     {D0 Collaboration}
\newc{\HXColl}     {H1 Collaboration}
\newc{\ZEUSColl}   {ZEUS Collaboration}
\newc{\LEPColl}    {LEP Collaboration}
\newc{\ATLASColl}  {ATLAS Collaboration}
\newc{\CMSColl}    {CMS Collaboration}
\newc{\UAColl}    {UA Collaboration}
\newc{\KAMLANDColl}{KamLAND Collaboration}
\newc{\IMBColl}    {IMB Collaboration}
\newc{\KAMIOColl}  {Kamiokande Collaboration}
\newc{\SKAMIOColl} {Super-Kamiokande Collaboration}
\newc{\SUDANTColl} {Soudan-2 Collaboration}
\newc{\MACROColl}  {MACRO Collaboration}
\newc{\GALLEXColl} {GALLEX Collaboration}
\newc{\GNOColl}    {GNO Collaboration}
\newc{\SAGEColl}  {SAGE Collaboration}
\newc{\SNOColl}  {SNO Collaboration}
\newc{\CHOOZColl}  {CHOOZ Collaboration}
\newc{\PDGColl}  {Particle Data Group Collaboration}

\def\issue(#1,#2,#3){{\bf #1}, #2 (#3)}
\def\ASTR(#1,#2,#3){Astropart.\ Phys. \issue(#1,#2,#3)}
\def\AJ(#1,#2,#3){Astrophysical.\ Jour. \issue(#1,#2,#3)}
\def\AJS(#1,#2,#3){Astrophys.\ J.\ Suppl. \issue(#1,#2,#3)}
\def\APP(#1,#2,#3){Acta.\ Phys.\ Pol. \issue(#1,#2,#3)}
\def\JCAP(#1,#2,#3){Journal\ XX. \issue(#1,#2,#3)} 
\def\SC(#1,#2,#3){Science \issue(#1,#2,#3)}
\def\PRD(#1,#2,#3){Phys.\ Rev.\ D \issue(#1,#2,#3)}
\def\PR(#1,#2,#3){Phys.\ Rev.\ \issue(#1,#2,#3)} 
\def\PRC(#1,#2,#3){Phys.\ Rev.\ C \issue(#1,#2,#3)}
\def\NPB(#1,#2,#3){Nucl.\ Phys.\ B \issue(#1,#2,#3)}
\def\NPPS(#1,#2,#3){Nucl.\ Phys.\ Proc. \ Suppl \issue(#1,#2,#3)}
\def\NJP(#1,#2,#3){New.\ J.\ Phys. \issue(#1,#2,#3)}
\def\JP(#1,#2,#3){J.\ Phys.\issue(#1,#2,#3)}
\def\PL(#1,#2,#3){Phys.\ Lett. \issue(#1,#2,#3)}
\def\PLB(#1,#2,#3){Phys.\ Lett.\ B  \issue(#1,#2,#3)}
\def\ZP(#1,#2,#3){Z.\ Phys. \issue(#1,#2,#3)}
\def\ZPC(#1,#2,#3){Z.\ Phys.\ C  \issue(#1,#2,#3)}
\def\PREP(#1,#2,#3){Phys.\ Rep. \issue(#1,#2,#3)}
\def\PRL(#1,#2,#3){Phys.\ Rev.\ Lett. \issue(#1,#2,#3)}
\def\MPL(#1,#2,#3){Mod.\ Phys.\ Lett. \issue(#1,#2,#3)}
\def\RMP(#1,#2,#3){Rev.\ Mod.\ Phys. \issue(#1,#2,#3)}
\def\SJNP(#1,#2,#3){Sov.\ J.\ Nucl.\ Phys. \issue(#1,#2,#3)}
\def\CPC(#1,#2,#3){Comp.\ Phys.\ Comm. \issue(#1,#2,#3)}
\def\IJMPA(#1,#2,#3){Int.\ J.\ Mod. \ Phys.\ A \issue(#1,#2,#3)}
\def\MPLA(#1,#2,#3){Mod.\ Phys.\ Lett.\ A \issue(#1,#2,#3)}
\def\PTP(#1,#2,#3){Prog.\ Theor.\ Phys. \issue(#1,#2,#3)}
\def\RMP(#1,#2,#3){Rev.\ Mod.\ Phys. \issue(#1,#2,#3)}
\def\NIMA(#1,#2,#3){Nucl.\ Instrum.\ Methods \ A \issue(#1,#2,#3)}
\def\JHEP(#1,#2,#3){J.\ High\ Energy\ Phys. \issue(#1,#2,#3)}
\def\EPJC(#1,#2,#3){Eur.\ Phys.\ J.\ C \issue(#1,#2,#3)}
\def\RPP (#1,#2,#3){Rept.\ Prog.\ Phys. \issue(#1,#2,#3)}
\def\PPNP(#1,#2,#3){ Prog.\ Part.\ Nucl.\ Phys. \issue(#1,#2,#3)}
\def\PS(#1,#2,#3){Phys.\ Scripta \issue(#1,#2,#3)}
\newc{\PRDR}[3]{{Phys. Rev. D} {\bf #1}, Rapid  Communications, #2 (#3)}
\def\PROP(#1,#2,#3){Prog.\ Part.\ Nucl.\ Phys. \issue(#1,#2,#3)}


\vspace*{\fill}
\vspace{-0.8in}
\begin{flushright}
{\tt IISER/HEP/01/09}
\end{flushright}

\begin{center}
{\Large \bf
Probing $R$-parity violating models of neutrino mass at the LHC via top
squark decays
}
\vglue 0.4cm
  Amitava Datta$^{(a)}$\footnote{adatta@iiserkol.ac.in} and
  Sujoy Poddar$^{(a)}$\footnote{sujoy$\_$phy@iiserkol.ac.in}
\vglue 0.1cm
 {\it $^{(a)}$
  Indian Institute of Science Education and Research, Kolkata, \\
  HC-VII, Sector III, Salt Lake City, Kolkata 700 106, India.
    \\}

\end{center}
\vspace{.2cm}

\begin{abstract}
{\noindent \normalsize}

It is shown that the $R$-parity violating decays of the lighter top squarks
($\lstop$)
triggered by the lepton number violating couplings $\lambda^{\prime}_{i33}$,
where the lepton family index i = 1-3,
can be observed at the LHC via the dilepton di-jet channel 
even if the coupling is as small as 10$^{-4}$ or
 10$^{-5}$, which is the case in several models of neutrino mass, provided
it is the next lightest supersymmetric particle(NLSP) the lightest neutralino
being the  lightest supersymmetric particle(LSP). We have first obtained a
fairly model independent estimate of  the
minimum observable value of the parameter 
($P_{ij} \equiv  BR(\widetilde t \ra l_i^+ b) \times BR(\widetilde t^* \ra l_j^- \bar b$)) at the LHC for an 
integrated luminosity of 10fb$^{-1}$ as a function of $\mlstop$ by a 
standard Pythia based analysis. We have then computed the parameter $P_{ij}$
in several representative models of neutrino mass constrained by the 
neutrino oscillation data and have found that the 
theoretical predictions are above the estimated minimum observable levels for
a wide region of the parameter space.    

\end{abstract}

\section{Introduction} \label{intro4}

 Neutrino oscillation experiments\cite{other}
 have confirmed that neutrinos indeed have very tiny masses, several orders 
 of magnitude smaller than any other fermion mass in the Standard Model (SM).
 The tiny masses of the neutrinos, however small, provide evidences of new 
 physics beyond the SM. 
 
 Neutrinos can be either Dirac fermions or Majorana fermions depending upon 
 whether the theory is lepton number conserving or violating. In the 
 SM, as originally proposed by Glashow, Salam and Weinberg, neutrinos are 
 massless since right handed neutrinos and lepton number violating terms are 
 not included.

 Both $R$-parity conserving (RPC) or $R$-parity violating (RPV)
 minimal supersymmetric 
 extension of the SM (MSSM)\cite{susyrev} are attractive examples of
 physics beyond the SM. In general the MSSM may contain RPC as well as
 RPV couplings. The latter include both lepton number and 
 baryon number violating terms which result in catastrophic proton decays.
 One escape route is to impose $R$-parity as a symmetry which eliminates all RPV
 couplings. This model is generally referred to as the RPC MSSM. However,
neutrino masses can be naturally introduced in this model only if it is embeded
in a grand unified theory (GUT) \cite{GUT}. Tiny Majorana neutrino masses are 
then generated by the see-saw mechanism \cite{seesaw}. Proton decay is a 
crucial test for most of the models belonging to this type.

 However, an attractive alternative for generating Majorana masses of the neutrinos
 without allowing proton decay is to impose a discrete symmetry which eliminates
 baryon number violating couplings from the RPV sector of the MSSM but retains 
 the lepton number violating ones. The observation of neutrinoless double beta decay\cite{double} and absence of
 proton decay  may be the hallmark of such RPV models of neutrino mass.
  
The GUT based models though very elegant have hardly any unambiguous prediction 
which may tested at the large hadron collider(LHC). 
In contrast the RPV models of neutrino mass are based
on TeV scale physics and, consequently, have many novel collider
signatures.

The observables in the neutrino sector not only 
depend on the RPV parameters but also on the RPC ones like the masses of the 
superpartners generically called sparticles. Thus the precise 
determination of the neutrino masses and mixing angles in neutrino 
oscillation experiments together with the measurement of sparticle masses and 
branching ratios (BRs) at collider experiments can indeed test the viability
of the RPV models quantitatively. Moreover the collider signatures of this
model are quite distinct from that of the RPC model. In this
paper our focus will be on a novel signature of a RPV model of $\nu$ mass
which can be easily probed  at the early stages of the upcoming LHC experiments.

In the RPC models the lightest supersymmetric particle (LSP) decays into lepton
number violating channels producing signals with high multiplicity but 
without much
missing energy which are in sharp contrast with the signals in a typical 
RPC model.
In RPV MSSM the sparticles other than the LSP can also directly
decay via lepton number violating channels which may lead to spectacular 
collider signatures. However, in a typical model of neutrino mass consistent
with the oscillation data such couplings turn out to be so small\cite{rpv} 
that the RPC decay of the sparticles overwhelm the RPV decays. Thus the LSP
decay is the only signature of $R$-parity violation.

However, the scenario changes dramatically if we consider the direct RPV 
decay of the lighter top squark ($\lstop$) 
\cite{nmass,biswarup,naba,shibu} with the assumption that $\lstop$ it is 
the next lightest supersymmetric particle(NLSP) while the lightest 
neutralino ($\lspone$) is the LSP. The theoretical motivation for the 
$\lstop$-NLSP scenario is the fact that it's superpartner - the top 
quark- is much heavier than any other matter particle in SM. This large 
top mass ($m_t$) leads to a spectacular mixing effect in the top squark 
mass matrix which suppresses the mass of the lighter eigenstate 
\cite{susyrev}. We assume that $\lstop$-NLSP decays via the loop 
induced mode $\lstop \ra c \lspone$ 
\cite{hikasa} and the four body\cite{boehm} decay mode, which occurs 
only in higher order of perturbation theory, with significant BR. 
The validity of this assumption will be justified later.
The RPV decays can now naturally compete with the RPC ones in 
spite of the fact that couplings underlying the former modes are highly 
suppressed by the $\nu$ oscillation data \cite{global}.

The lighter top squark decays  into a lepton and a $b$-jet via RPV couplings
$\lambda_{i33} '$ are listed below:
\be
a)~~\lstop \rightarrow l_i^+ b~ ;~~~~b)~~\lstopbar \rightarrow l_i^- \bar b
\ee
where $i$=1-3 corresponds to $e$, $\mu$ and $\tau$ respectively. Our 
signal consists of opposite sign dileptons(OSDL), two hard jets with 
very little $\met$. These modes dominate, e.g., in many RPV models where 
neutrino masses are generated at the one loop level by the 
$\lambda'_{i33}$ couplings, where i is the lepton index and 3 stands for 
quarks or squarks belonging to the third generation (see below).

We take the lowest order QCD cross section of top squark pair production 
which depends on $\mlstop$ only. Requiring that 
the significance of the signal over the SM background be at least 5 
$\sigma$ level for an integrated luminosity of 10 $fb^{-1}$ or smaller, 
we can then put fairly model independent lower limits on the products of the 
BRs (PBRs) of the RPV decay modes in Eq. 1. In our analysis both 
the signal and the backgrounds are simulated with Pythia. As expected 
the  range of $\mlstop$ which can be probed at the LHC is 
significantly 
larger compared to the reach of Tevatron RUN II\cite{cdf,admgspd}. The 
details of our simulations will be presented in the next section.

In principle the viability of probing any RPV model of neutrino mass 
with the above characteristics at the LHC
can be checked by computing PBRs in respective models, and comparing 
with 
the estimated lower limits. For the purpose of illustration we have 
considered 
in section 3 a model based on three bilinear RPV couplings ($\mu_i$) and 
three trilinear couplings ($\lambda'_{i33}$) at the weak scale  
\cite{abada} and have carried out the above check. 
It is gratifying to note that most of the parameter space allowed
by the neutrino oscillation data can be probed by the early LHC 
experiments with an integrated luminosity of 10 fb$^{-1}$ (see section 3). 
Moreover, the constraints 
from oscillation data indicate that the $\lambda'_{i33}$ couplings 
should have certain hierarchical pattern leading to distinct collider 
signatures \cite{adspdsp} . This hierarchy among the couplings can be 
qualitatively tested by observing the relative sizes of signals involving 
different OSDL signals.

The summary, the conclusions and future outlooks are in the last section.

\section{ The signals and the SM backgrounds} \label{result}

The production and decay of the lighter top squark pairs are simulated by
Pythia\cite{pythia}. Initial and final state radiation,
decay, hadronization, fragmentation and jet formation are implemented
following the standard procedures in Pythia. We have considered only
the RPV decay modes of $\lstop$ via the couplings 
$\lambda '_{i33}$ ,i = 1-3 (Eq. 1) and in this section
their BRs are taken to be free parameters.
We have used the toy
calorimeter simulation (PYCELL) in Pythia with the following
criteria:

\begin{itemize}

\item The calorimeter coverage is $\vert \eta \vert < 4.5$. The segmentation is given by $\Delta \eta \times \Delta \phi = 0.09 \times 0.09$ which resembles a generic LHC detector.

\item A cone algorithm with $\Delta R(j,j)$ = $\sqrt {\Delta\eta^2 + 
\Delta\phi^2}= 0.5 $ has been used for jet finding.

\item Jets are ordered in 
E$\mathrm{_T}$ and E$^{\mathrm{jet}}_{\mathrm{T,min}} = 30 $GeV.
\end{itemize}

Various combinations of OSLDs in the final state are selected as follows:

\begin{itemize}
\item Only tau leptons decaying into hadrons are selected 
provided the resulting jet has 
P$\mathrm{_T \ge 30}$ GeV  and $\vert\eta \vert < 3.0$.


\item Leptons $(l=e,\mu)$ are selected with P$\mathrm{_T \ge 20}$ GeV
and $\vert\eta \vert < 2.5$. 

 \end{itemize}

The following selection criteria(SC) are used for background rejection :
\begin{itemize}
\item The $\tau$-jets are tagged according to the tagging 
efficiencies provided by the CMS collaboration\cite{cms}(Fig. 12.9)(SC1).
Hadronic BR of the $\tau$ is also included in the corresponding 
efficiency. For $e$ and $\mu$ SC1 is the lepton-jet
isolation cut.
We require $\Delta R(l,j) > 0.5$.The detection efficiency of the leptons 
are assumed to be approximately $ 100 \%$ for simplicity.

\item Events with two 
isolated leptonic objects (e,$\mu$ or tagged $\tau$-jets ) 
are rejected if P$\mathrm{_T \le 150}$ GeV, where $l$ = $e$ or $\mu$ (SC2)
or $E^{V(\tau)}_T$ $< $100 GeV, where $E^{V(\tau)}_T$ is the $E_T$ of the $\tau$
jet.

\item We select events with exactly two jets other than the tagged 
$\tau$-jets (SC3). The event is rejected if the additional jets have
 P$\mathrm{_T \le 100}$ GeV (SC 4). \footnote{It is expected that this cut 
 would also suppress the SUSY backgrounds due to, e.g., $\tilde q, \tilde g$
 production.} 

\item Events with missing transverse energy ($\etslash) > 60$ GeV are rejected 
(SC5).

\end{itemize}

Through SC1 we have  severely constrained the transverse momentum of two 
leptons $l=e,\mu$ to reject the leptons coming from the leptonic decays 
of the tau. Moreover such a strong cut reduces most of the SM 
backgrounds significantly. We have considered 
backgrounds from:
 $ W W,  W Z,  Z Z,  t \bar t$, Drell-Yan (DY) and QCD events.
The missing energy veto plays a crucial role to tame down 
$W W$ and $t \bar t$ backgrounds  as they are rich in missing energy. 
 Mistagging of light jets as $\tau$-jets
is a major source of background to di-tau events. We have taken this 
into
account. However, if we also  employ $b$-tagging  then this  
background can be brought under control to some extent.  

In our work  $b$ tagging has been implemented  according to the
following prescription. A jet with $\vert \eta \vert < 2.5$
matching with a B-hadron of decay length $ > 0.9 ~\mathrm{mm} $ has
been marked $ tagged$. The above criteria ensures that $\epsilon _{b}
\simeq 0.5$ in $ t \bar t$ events, where $\epsilon _{b}$ is the
 single $b$-jet tagging efficiency (i.e., the ratio of the number of tagged
 $b$-jets and the number of taggable $b$-jets in $t \bar t$ events).

The leading order(LO) cross-sections for $\lstop-\lstop^*$ pairs 
presented in Table 1 are computed using calcHEP (version 
2.3.7)\cite{calchep}. 

\begin{table}[!ht]
\begin{center}\

\begin{tabular}{|c|c|c|c|c|c|}
       \hline

Signal &240&300&400&450&500\\
\hline
$\sigma(pb)$ & 14.6 &4.8&1.1&0.58&0.32\\

\hline

\end{tabular}
\end{center}
   \caption{ $\lstop$ - $\lstop^*$ pair production cross section 
($\sigma$) at the LHC 
for  different $\mlstop$.}

\end{table}

In Table 2 we have presented the combined  efficiencies of SC1 - SC5
in steps.
 The first column of Table 2 shows
signals with different topology of final states. Here $e~e~X$, $\tau
\tau X$, $e \tau X$ and $ e \mu X$ represent final states 
without b-jet tagging. The cumulative efficiency of each SC
 for $\mlstop$ = 400 GeV is presented in the next five columns. However, 
we have not separately presented the efficiencies corresponding to final 
states with muons as we have assumed that both $e$ and $\mu$ are 
detected with approximately 100$\%$ efficiency.

Table 3 contains the effect of b-jet tagging on different final
states.
We have used the notations $0b$, $1b$ and
$2b$ to specify  signal events  with zero, one  and
two  tagged b-jets respectively. 
From this Table it is also evident that the efficiencies increase for
larger $\mlstop$ since  the $P_T$ cut on leptons become
less severe. This compensates the fall of the cross section with
increasing $\mlstop$ to some extent.

\begin{table}[!ht]
\begin{center}

\begin{tabular}{|c|c|c|c|c|c|}
\hline

$\lstop \lstopbar$ &$\epsilon_1$&$\epsilon_2$&$\epsilon_3$&$
\epsilon_4$& $\epsilon_5$\\
\hline
$e e X$ &0.93708   &0.292239 &  0.087228 &
0.043344& 0.032823
\\
\hline
\hline
$\tau \tau X$ &0.251343 & 0.111546 & 0.033201 &0.031554 &0.008955 \\
\hline
\hline
$e \mu X$ & 0.94101& 0.295239 & 0.088216& 0.043415& 0.033060\\
\hline
\hline
$e \tau X$ & 0.474948& 0.180945 & 0.053820& 0.044793& 0.016965\\
\hline
\end{tabular}
\end{center}

\caption{Efficiency table for  $\mlstop =400 \gev.$}

\end{table}


\begin{table}[!ht]
\begin{center}\

\begin{tabular}{|c|c|c|c|c|c|}
       \hline
&\multicolumn{5}{c|}{SIGNAL}\\

\hline
$\mlstop(\gev)$ &240&300&400&450&500\\
\hline
$e~ e~ 0b$          &0.00032&0.00066 &0.00189&0.00234&0.00255 \\
\hline
$e~ e~ 1b$          &0.00121& 0.00330&0.01116&0.01461&0.01580\\
\hline
$e~ e~ 2b$          &0.00176& 0.00509& 0.01984&0.02620&0.03121\\
\hline
$ e~ e ~X$            &0.00328&0.00905&0.03282&0.04315&0.04957\\
\hline
$\tau~ \tau~ 0b$    &0.00059&0.00073 &0.00112&0.00091&0.00097 \\
\hline
$\tau~ \tau~ 1b$    &0.00153&0.00284&0.00363&0.00351&0.00391\\
\hline
$\tau~ \tau~ 2b$    &0.00126 &0.00226&0.00421&0.00450&0.00522\\
\hline
$ \tau~ \tau ~X$      &0.00338&0.00582&0.00896& 0.00892&0.01098\\
\hline
$\tau~ e~ 0b$       &0.00045&0.00081 &0.00148& 0.00142& 0.00142\\
\hline
$\tau~ e~ 1b$       &0.00135&0.00307&0.00667& 0.00705&0.00717\\
\hline
$\tau~ e~ 2b$       &0.00126 & 0.00346&0.00882 &0.00997&0.01078\\
\hline
$ \tau~ e ~X$         & 0.00308& 0.00734&0.01697&0.01843&0.01936\\
\hline
$\mu~ e~ 0b$        &0.000315 &0.00067&0.00190&0.00235&0.00257\\
\hline
$\mu~ e~ 1b$        &0.00123 &0.00334&0.01125&0.01469&0.01635\\
\hline
$\mu~ e~ 2b$        &0.00178 &0.00512&0.01992&0.02625&0.03129\\
\hline
$ \mu~ e ~X$          &0.00332 &0.00912&0.03306&0.04329&0.05021\\
\hline

\end{tabular}
\end{center}
   \caption{Final efficiencies for different $\mlstop$ (including 
b-tagging if implemented).
}
\end{table}





\begin{table}[!ht]
\begin{center}

\begin{tabular}{|c|c|c|c|c|c|}
\hline

$t \bar t$ &$\epsilon_1$&$\epsilon_2$&$\epsilon_3$&$
\epsilon_4$& $\epsilon_5$\\
\hline
$e e$ &$7.63 \times 10^{-3}$   &$2.22\times 10^{-5}$ & $ 5.70 \times 10^{-6}$ & 
$7.00 \times 10^{-7}$&$1.00 \times 10^{-7}$
\\
\hline
\hline
$\tau \tau$ &$4.76\times 10^{-4}$ &$4.00\times 10^{-6}$ &$1.50 \times 10^{-6}$ &
$1.00\times 10^{-6}$& $4.00 \times 10^{-7}$ \\
\hline
\hline
$e \mu$ &$7.74 \times 10^{-3}$  &$2.01 \times 10^{-5}$ &$6.01 \times 10^{-6}$ &
$6.80 \times 10^{-7}$ & $5.0 \times 10^{-7}$\\
\hline
\hline
$e \tau$ &$1.88 \times 10^{-3}$ &$9.3\times 10^{-6}$ & $2.95 \times 10^{-5}$  &$ 9.50\times 10^{-7}$ & $2.00 \times 10^{-7}$\\
\hline

\end{tabular}
\end{center}

\caption{Efficiency table for  $t \bar t$ process}

\end{table}

In Table 4 we have shown the effect of cuts on the background 
from $t \bar t$ events. 
SC2 is very effective  in reducing this  background significantly. 
Moreover this background is accompanied by large amount 
of $\met$ and SC5  also reduces it significantly. 
Since $t \bar t$ decays contain two $b$ quarks,  $b$- tagging is not 
very effective here and has not been included in Table 4.

\begin{table}[!ht]
\begin{center}

\begin{tabular}{|c|c|c|c|c|c|}
\hline

$QCD$ &$\epsilon_1$&$\epsilon_2$&$\epsilon_3$&$
\epsilon_4$& $\epsilon_5$\\
\hline
$e e$ &$1.16 \times 10^{-5}$   &0 & 0 & 0 & 0 \\
\hline
\hline

$\tau \tau$ &$9.10 \times 10^{-3}$   & $4.02 \times 10^{-3} $ & $1.05 
\times 10^{-3}$ & $2.85 \times 10^{-4} $ & $2.10 \times 10^{-4}$ \\
\hline
\hline

$e \mu$ &$6.0 \times 10^{-6}$   &0 & 0 & 0 & 0 \\
\hline
\hline

$ e\tau$ &0   &0 & 0 & 0 & 0 \\
\hline

\end{tabular}
\end{center}

\caption{Efficiency table for the $QCD$ process in the $ \hat p_T$ bin:
 400 GeV $< \hat p_T <$ 1000 GeV.}

\end{table}

 Table 5 presents another important  background arising from the $2 \ra 
2$ processes due to pure
$QCD$ interactions for 400 GeV $< \hat p_T <$ 1000 GeV, 
where $\hat p_T$ is the transverse  momentum of the two partons in the 
final state . However, SC2 
 completely kills  all backgounds except for those with the di-$\tau$  
final states. The latter 
 background, mainly
due to  mistagging of light flavour jets as $\tau$-jets, affect
the di-$\tau$ signal very seriously . The 
mistagging
probability has also been taken from \cite{cms} (Fig. 12.9).

This background
is  very large, as expected, since the $QCD$ cross-section is very 
large. 
The leading order  cross-sections have been computed by Pythia in two 
$ \hat p_T$ bins :
(i) 400 GeV $< \hat p_T <$ 1000 GeV   and
(ii) 1000 GeV$< \hat p_T <$ 2000 GeV .
 We have chosen the QCD scale to be $\sqrt{\hat s}$. 
The corresponding cross-sections
being 2090$\pb$ and 10$\pb$ respectively. Beyond 2000 GeV
the number of events are negligible. We shall discuss later how
the visibility of the di-$\tau$ signal can be improved by
employing $b$ tagging.



\begin{table}[!ht]
\begin{center}\

\begin{tabular}{|c|c|c|c|c|c|c|}
       \hline
&\multicolumn{6}{c|}{BACKGROUND}\\

       \hline

Final state &$W^+W^-$&$W^{\pm}Z$&$ZZ$&$t \bar t$& QCD& DY\\
\hline
\hline
$\sigma(pb)$ & 73.5 &33.4&10.1&400&2090,10.6&3400\\

\hline
\hline

$e~ e~ 0b$ &0.37&0.33& 0.40&0.40&-&-\\
\hline
$e~ e~ 1b$ &-&-&-&- &-&-\\
\hline
$e~ e~ 2b$ &-&-&-&-&-&-\\
\hline
\hline
$ e~ e $ &0.37&0.33&0.40&0.40&-&-\\

\hline
\hline
$\tau~ \tau~ 0b$ &-&-&-&-&4218&- \\
\hline
$\tau~ \tau~ 1b$ &-&-&-&0.80&143&-\\
\hline
$\tau~ \tau~ 2b$ &-&- &0.20&0.80&12&-\\
\hline
\hline
$ \tau~ \tau $ &-&- & 0.20&1.60& 4373&-\\
\hline
\hline
$\tau~ e~ 0b$ &-&-&-&-&-&- \\
\hline
$\tau~ e~ 1b$ &- &-&-&0.40&-&-\\
\hline
$\tau~ e~ 2b$ &- &-&-&0.40&-&-\\
\hline
\hline
$ \tau~ e $ &- &-&-&0.80&-&-\\
\hline
\hline
$\mu~ e~ 0b$ &0.37 &-&-&0.40&-&-\\
\hline
$\mu~ e~ 1b$ &- &-&-&0.80&-&-\\
\hline
$\mu~ e~ 2b$ &- &-&-&0.80&-&-\\
\hline
\hline
$\mu~ e $ &0.37 &-&-&2.00&-&-\\

\hline
\hline
\end{tabular}
\end{center}
 \caption{ Total number of all types of backgrounds survived after all cuts
 .}

\end{table}

In Table 6 we have computed the  numerically significant 
backgrounds  of all types for 
$\lum$ = 10 $\ifb$. Here '-' denotes a vanishingly small background.
It is clear from this  table that only $t \bar t$ and
QCD backgrounds are relevant.  
The LO cross-sections in the second row of Table 6 except for the QCD 
processes have been computed 
using calcHEP(version 2.3.7)\cite{calchep}. 
Due to very strong cut on $P_T$ of highest two leptons SC2 DY 
type 
backgrounds become vanishingly small. Moreover, SC3 and SC4 finally 
reduce it to zero. Other backgrounds like $WW$, $WZ$ and $ZZ$ become 
vanishingly small mainly due to SC2.

The Product Branching Ratio (PBR) is defined as :
\be
P_{ij} \equiv  BR(\lstop \ra l_i^+ b) \times BR(\lstop^* \ra l_j^- \bar b)
\ee
where $i$ or $j$ can run from 1-3 corresponding to $e$, $\mu$ and 
$\tau$ respectively. The  Minimum Observable Product Branching 
Ratio(MOPBR $\equiv P_{ij}^{min}$) corresponds
to $S/\sqrt(B) \geq 5$, where $S$ and $B$ are the number of
signal and background events 
respectively. However, for a typical signal with negligible  background 
we have 
required  $S \geq 10$ as the limit of observability and MOPBR is 
computed accordingly.

For a given $\lum$ the MOPBR for each process is  computed from Table 3
and Table 6 by following expression:\\
 
\begin{eqnarray}
P_{ij}^{min}& = &5 \sqrt {\eta\lum \Sigma \sigma^b \eps^b }
\over \eta \lum \sigma(\lstop \lstopbar) \eps,
\end{eqnarray}
\noindent
where $P_{ij}$ is already defined in Eq. 2.
$\sigma^b$ and  $\eps^b$ (not to be  confused with  $\eps_b$, the
$b$-jet tagging efficiency) denote the cross section
and  the efficiency  of  background of type $b$ .
Similarly  $\eps$ is the final efficiency for the signal.
$\eta$ is 2 for $i \neq j$ and $\eta$ is 1 for $i = j$.
The integrated luminosity $\lum$ is taken to be
10 fb$^{-1}$. The estimated MOPBRs  are given in Table 7(without
$b$-jet tagging) and  Table 8 ( with two tagged $b$-jets).

We remind the reader that in Table 7 and Table 8 a signal is assumed to 
be observable if S $\geq$ 10 even if B is $\leq$ 4. In Table 7 and Table 
8 a '$\times$' indicates that corresponding channel can not be probed.

Our conclusions  so far have been based on LO cross sections. If the
next to leading order corrections are included the $\lstop -
\lstop^*$ production cross section is enhanced by 30 - 40 \% due
to a K-factor\cite{NLO}. It is then clear from Eq. 3, that the
estimated MOPBR would remain unaltered even if all significant 
background cross sections are enhanced by a factor of two due to
higher order corrections. 

\begin{table}[!ht]
\begin{center}\

\begin{tabular}{|c|c|c|c|c|c|}
\hline

$\mlstop(\gev)$ &240&300&400&450&500\\
\hline
$P_{11}^{min}(\%)$    & 2.1  & 2.3 & 2.8 & 4.0 & 6.3\\
\hline
$P_{33}^{min}(\%)$    & $\times$     &$\times$     &$\times$     &$\times$     &$\times$   \\
\hline
$P_{12}^{min}(\%)$    & 1.0   & 1.1  & 1.4  & 2.0  & 3.4 \\
\hline
$P_{13}^{min}(\%)$    & 1.1   & 1.4  & 2.7  & 4.7  & 8.0 \\
\hline
\end{tabular}
\end{center}
\caption{Minimum value of PBR estimated from the sample without $b$ tagging
 .}
\end{table}


\begin{table}[!ht]
\begin{center}\

\begin{tabular}{|c|c|c|c|c|c|}
\hline

$\mlstop(\gev)$ &240&300&400&450&500\\
\hline
$P_{11}^{min}(\%)$    & 3.9  & 4.1 & 4.5 &6.6 &10.0 \\
\hline
$P_{33}^{min}(\%)$    & 9.4 &16.0 &37.5  & 66.4& $\times$   \\
\hline
$P_{12}^{min}(\%)$    & 1.9  &2.0  &2.3   &3.3  &5.0  \\
\hline
$P_{13}^{min}(\%)$    & 2.7  &3.0  &5.2   &8.6  &14.5 \\
\hline

\end{tabular}
\end{center}
\caption{ Minimum value of PBR estimated from the 2-$b$ tagged sample.}
\end{table}


\begin{figure}[!htb]
\begin{center}
\hspace*{-1.0cm} \mbox{\epsfxsize=.5\textwidth
                 \epsffile{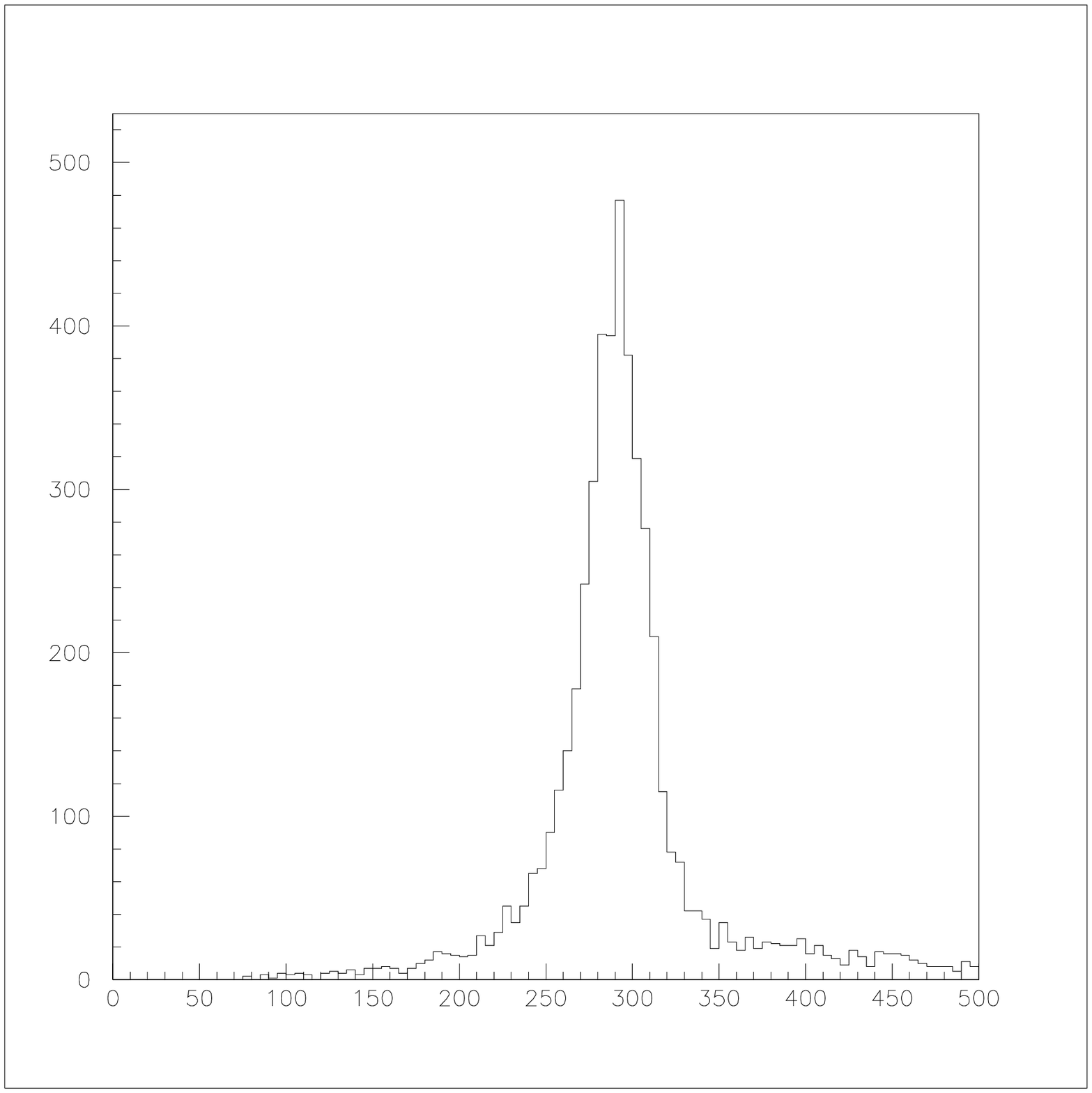}}
\caption{The invariant mass distribution for $\mlstop = 300 \gev$.}
\end{center}
\end{figure}


We present in Fig. 1 the distribution (unnormalised) of invariant mass
of a electron -jet pair in the dielectron-dijet sample without b-tagging 
for $\mlstop=300\gev$. We first reconstruct invariant 
mass for all possible electron-jet pair. Among these pairs We  select the 
two such that the difference in their invariant mass is minimum. We then
plot the higher of the two invariant masses. This peak, if observed, would
unambiguously establish the lepton number violating nature
of the underlying decay. In contrast if neutralino decay is the only signal
of $R$-parity violation, then this information may not be available. 
For example, if $\lspone \rightarrow \nu b \bar b$ is the dominant
decay mode of the LSP via the $\lambda^{\prime}_{i33}$ coupling then the 
lepton number violating nature of the decay dynamics will be hard to 
establish.


In the next  section we shall calculate the PBR for different signals in 
a realistic models of neutrino
mass constrained by the neutrino oscillation data
and examine whether the  predictions  exceed the corresponding MOPBR
estimated in this section. Our main aim is to illustrate that the LHC 
experiments will be sufficiently sensitive to probe these  models and not to 
make an exhaustive study of all possible models.

\section{Model Calculations}

The collider signatures considered in the last section arise only in
models with non-vanishing trilinear $\lambda'_{i33}$ type couplings
at the weak scale.
However, consistency with neutrino oscillation data require the
introduction of more  RPV parameters (bilinear superpotential terms,
bilinear soft breaking terms etc)\cite{subhendu}. In fact the list of possible choices
is quite long. It is expected that the
constraints on the $\lambda'$ couplings in the most general  model
imposed by the $\nu$ - oscillation data will
be considerably weaker  and the observability of the resulting dilepton-dijet 
signal will improve.  
 Thus we have
restricted ourselves to models with a minimal set  of parameters capable
of explaining the oscillation data with rather stringent constraints on
the $\lambda'$ couplings.

 We work in a
 basis where the sneutrino vevs are zero. It is assumed that in this basis only
 three nonzero bilinear($\mu_i$) and three trilinear($\lbp_{i33}$) couplings,
 all
 defined at the weak scale, are numerically significant. In this framework the
 neutrino mass matrix receives contributions both at the tree and one loop 
 level.
 It should be emphasised that the tree level mass matrix,
 which is
 independent of $\lambda_{i33}$ couplings,
 yields only two massless 
 neutrinos. Thus the interplay of the tree level and one loop mass matrices is
 essential for consistency with the oscillation data.

The chargino-charge lepton, the neutralino -
neutrino and other relevant mixing matrices in this basis may be found in
\cite{subhendu}. In principle the diagonalization of these matrices may
induce additional lepton number violating couplings
which can affect the BRs of the top squark decays
considered in this paper.  For example, the RPC coupling $\lstop - t - \widetilde W_3$
may induce new RPV vertices through $\widetilde W_3 - \nu$ mixing. However, it was 
shown in\cite{adsp} that the new modes induced in this way would have negligible BRs.
As a result the approximation that the decays
of the top squark NLSP  are driven by the $\lbp_{i33}$
couplings only is justified.

In addition to the RPV parameters the neutrino masses and mixing
angles depends on RPC parameters. In this paper we shall
use the following popular assumptions to reduce the number of free
parameters in the RPC sector: i) At the weak scale the soft breaking mass
squared parameters of the L and R-type squarks belonging to the third 
generation are assumed to be the same( the other squark masses are not 
relevant
for computing neutrino masses and mixing angles in this model).
ii) We shall also use the relation $M_2 \approx 2~ M_1$ at the weak
scale as is the case in models with a unified gaugino mass at $M_G$.
Here $M_1$ and $M_2$ are respectively the soft breaking masses
of the U(1) and SU(2) gauginos respectively.

The tree level neutrino mass matrix and, hence, the predicted
neutrino masses  depends on the parameters of the gaugino
sector(through the parameter $C$\cite{abada,adspdsp}). They are $M_2$, $M_1$
, $\mu$ (the higgsino mass parameter) and tan $\beta=v_2/v_1$,
where $v_1$ and $v_2$ are the vacuum expectation values (vevs) for the down
type and the up type neutral higgs bosons respectively. We remind the
reader that
for relatively large tan $\beta$s the loop decay overwhelms the RPV decay
\cite{shibu,boehm}.
We have, therefore, restricted ourselves to $tan \beta$ = 5-8.
It is also
convenient to classify various models of the RPC sector according to the
relative magnitude of $M_2$ and $\mu$. If $M_1 < M_2$ $\ll$ $\mu$, then
the lighter chargino ($\tilde\chi_1^{\pm}$), the LSP
($\tilde\chi_1^0$) and the second lightest neutralino ($\tilde\chi_2^0$)
are dominantly gauginos. Such  models are referred to as the gaugino-like
model. On the other hand in the mixed model ($M_1$$ < M_2$$\approx$ $\mu$),
 $\tilde\chi_1^{\pm}$ and $\tilde\chi_2^0$ are admixtures of gauginos
and higgsinos. In both the cases, however, $\lspone$ is 
 purely a bino to a very good approximation.
There are models with $M_1$,$M_2$ $\gg$ $\mu$
 in which $\tilde\chi_1^{\pm}$,
 $\tilde\chi_1^0$ and $ \tilde\chi_2^0$  are higgsino -
 like and all have approximately the same mass ($\approx \mu$).
 It is difficult to accommodate the top squark NLSP
 in such models  without fine adjustments of the parameters. Thus the LSP
 decay seems to  be the only viable collider signature.
 One can also construct   models  wino or higgsino dominated
 LSPs. However, the $\lstop$-NLSP scenario cannot be naturally accommodated in
 these frameworks for reasons similar to the one in  the last paragraph.

 The one loop mass matrix, on the other hand, depends on the sbottom
 sector (through the parameter $K_2$
  \cite{abada,adspdsp}). This parameter decreases for
  higher values of the common squark mass for the third generation. From the
  structure of the mass matrix it then appears that for fixed C, identical
  neutrino masses and mixing angles can be obtained for higher values of the
  trilinear couplings if $K_2$ is decreased. Thus at the first sight it
  seems that arbitrarily large width of the RPV decays may be accommodated
  for any given neutrino data. This, however, is not correct because of the
  complicated dependence of the RPV and loop decay BRs of $\lstop$ on the
  RPC parameters and certain theoretical constraints. The common squark mass
  cannot be increased arbitrarily without violating the top squark NLSP
  condition. Of course larger values of the trilinear soft breaking term
  $A_t$ may restore the NLSP condition. But larger values of $A_t$ tend to
  develop a charge colour breaking( CCB ) minimum of the scalar potential
   \cite{ccb}. Finally the pseudo scalar higgs mass parameter $M_A$ can be
   increased to satisfy the CCB condition. But as noted earlier \cite{adspdsp}
   that would enhance the loop decay width as well and suppress the BRs of
   the RPV decay modes.

We have chosen the following
RPC scenarios : A) The gaugino dominated model  
 and B)The mixed type model. 
     The choice of
    RPC parameters for model A) and model B) are:\\
    A) $M_1=195.0,~ M_2= 370.0,~ \mu=710.0,~ \tan\beta=6.0,~ A_t=1100.0, 
    ~A_b=1000.0,~ M_{\tilde q}$(common squark mass )=$450.0$,~ $M_{\tilde l}$ (common slepton 
    mass ) = $400.0$ and  $M_A=500.0$ and 
    B) $M_1=170.0,~ M_2= 330.0,~ \mu=320.0,~ \tan\beta=6.0,~ A_t=1045.0,~ A_b=1000.0,~ M_{\tilde q}$=$450.0$, $M_{\tilde l}$ = $400.0$
    and  $M_A=200.0$, where all masses and mass
    parameters are in $\gev$. Both the scenarios correspond to 
$\mlstop=240 \gev$ and $\lstop$ is the NLSP.  It should be noted that the
 slepton mass is specified to ensure that the $\lstop$ is the NLSP. It
 does not affect the neutrino mass matrix. 

Even if $\lstop$ is the NLSP the following modes may compete with the 
RPV decays and overwhelm it: \\ 
\be
a)~~\lstop \rightarrow t \lspone ;~~~~b)~~\lstop \rightarrow b W \lspone
~~~c) \lstop \rightarrow c \lspone; ~~~d)~~ \lstop \rightarrow f \bar f b \lspone
\ee
In the parameter spaces we have worked with the mode a) is kinematically 
disallowed. The second mode is highly suppressed if the LSP is Bino dominated 
as is assumed in this analysis. Thus in the scenario under consideration only 
modes c) and d) may compete with RPV decays of $\lstop$. In this section we 
have computed the PBRs taking into account the competition among the above 
three modes. 

   Next  we have randomly generated bilinear and trilinear RPV 
   couplings, 
   $\mu_i$ and $\lambda '_{i33}$. Then these parameters are constrained 
   by $\nu$ oscillation data which allows very few sets of RPV parameters 
   for the above  RPC parameters. Most of the allowed trilinear RPV 
   couplings 
   lie within $10^{-4} - 10^{-5}$. Finally the relevant PBRs have been
   calculated in model (A) and (B).
   
 In Table 9 we  present several representative  sets  of trilinear RPV
  parameters allowed by $\nu$ oscillation data and the corresponding
 PBR. A `-' indicates that the predicted PBR is negligible.
 As noted before
two of the couplings turns out to be large while the third one is
suppressed due to oscillation constraints. 
It turns out 
that the PBR's involving the large couplings are larger than the 
corresponding MOPBRs estimated in 
the last section even without b-tagging(see Table 7). The only exception
is $P_{33}$ which cannot be probed without b-jet tagging (see Table 8)



\begin{table}[!ht]
\begin{center}\
\begin{tabular}{|c|c|c|c|c|c|c|c|c|}

\hline
${\lambda^{\prime}}_{133} [\times 10^{-5}]$&${\lambda^{\prime}}_{233} [\times 10^{-5}]$& ${\lambda^{\prime}}_{333} [\times 10^{-5}]$ &$P_{11}$&
$P_{22}$& $P_{33}$&$P_{12}$&$P_{23}$&$P_{13}$ \\
\hline
Model A\\
\hline
1.6 &8.3  & 10.0 & - & 5.1 & 10.8 & 0.2 & 7.4 & 0.3\\
\hline
7.5 & 0.7 & 9.2 & 4.9 & - & 11.4 & - & 0.1 & 7.5\\
\hline
4.6 & 4.5 & 0.3 & 6.8 & 6.8 & - & 6.7 & - & - \\

\hline

Model B\\
\hline
 11.9  & 0.99& 15.0&4.2 &- &10.6 &-& -&6.6\\
\hline
 0.59  & 13.6 & 16.8 &  -   & 4.3 & 10.2 & -  & 6.6 &-\\
\hline
 7.3  & 7.4& 0.9&6.3 &6.6 &- &6.4& 0.1&0.1\\

\hline 
\end{tabular}
\end{center}
\caption{ Trilinear RPV couplings allowed by $\nu$ oscillation data and the 
corresponding PBRs computed in models A and B (see text) with $\mlstop = 240 \gev
$. }
\end{table}


 For larger $\mlstop$, there exists allowed
 RPV parameter space with observable PBRs at the early LHC runs.
 However, if we go beyond $\mlstop=500 \gev$ the
di-tau channel cannot be probed even with $b$ -tagging.  
Nevertheless, observation of the $e-\tau$ and the $\mu -\tau$ channel
will provide  evidence for a relatively large $\lambda_{333}$. 
We present
in Table 10 for $\mlstop=500 \gev$. The  RPC parameters
corresponding to a  Gaugino model are chosen to be:\\
$M_1=475.0,~ M_2= 860.0, ~\mu=1650.0,~ \tan\beta=6.0,~ A_t=995.0,~ A_b=1000.0,
~M_{\tilde q}$=$575.0$, $M_{\tilde l}$ 
= $525.0$ and  $M_A=300.0$,
where all masses and mass parameters are in $\gev$. 

\begin{table}[!ht]
\begin{center}\
\begin{tabular}{|c|c|c|c|c|c|c|c|c|}

\hline
${\lambda^{\prime}}_{133} [\times 10^{-5}]$&${\lambda^{\prime}}_{233} [\times 10^{-5}]$& ${\lambda^{\prime}}_{333} [\times 10^{-5}]$ &$P_{11}$&
$P_{22}$& $P_{33}$&$P_{12}$&$P_{23}$&$P_{13}$ \\

\hline 
9.1 & 4.0 & 6.4 & 20.7 & - & 5.1 & 4.0 & 2.0 & 10.3\\
\hline 
4.4 & 10.9 & 5.6 & - & 31.4 & 2.2 & 5.1 & 8.3 & 1.3\\
\hline 
\end{tabular}
\end{center}
\caption{Same as Table 9 for $\mlstop = 500 \gev$.}
\end{table}
\noindent
We have checked that even for $\mlstop > 500 \gev$ there exits 
RPV parameter 
space allowed by oscillation data which leads to observable dilepton
-dijet signals in early LHC experiments.


\begin{table}[!ht]
\begin{center}\
\begin{tabular}{|c|c|c|c|c|c|}
\hline

\multicolumn{2}{|c|}{$(\lambda^{\prime}_{133})^{max}$} &\multicolumn{2}{c|}{$(\lambda^{\prime}_{233})^{max}$ } & \multicolumn{2}{c|}{ $(\lambda^{\prime}_{333})^{max}$}\\
\hline

\multicolumn{2}{|c|}{92} &\multicolumn{2}{c|}{2176 } & \multicolumn{2}{c|}{ 2086}\\
\hline
\cline{1-6}
$P_{11} , P_{13}$ & 56 & $P_{22} , P_{23}$ & 1376 & $P_{33} , P_{23}$ & 874\\
\hline
$P_{11} , P_{12}$ & 15 & $P_{22} , P_{12}$ &119 & $P_{33} , P_{13}$ & 27\\
\hline
$P_{11} $ & 2 & $P_{22} $ &664 & $P_{22} , P_{23}$ & 274\\
\hline
$P_{12} $ & 10 & $ * *$ &17 & $ P_{11} , P_{13}$ & 25\\
\hline
$ * *$ & 9 &  & & $P_{23} $ & 45\\
\hline
 &  &  & & $ P_{33} $ & 304\\
\hline
 &  &  & & $ * * $ & 537\\
\hline

\end{tabular}
\end{center}
   \caption{Number of allowed solutions in the mixed model($\mlstop=240 GeV$)
   consistent with $\nu$ oscillation data 
   which satisfy the MOPBR given in
   Table 7 and Table 8. The above numbers are estimated for $\lum = 10 \ifb$.
}
\end{table}
 
 We have randomly generated $10^9$ sets of RPV parameters in the 
mixed model with $\mlstop = 240 GeV$. Out of these only 4354 are 
consistent with the $\nu$-oscillation data. 
These solutions can be further  
classified into three groups according to the highest value of  $\lambda 
^{\prime}_{i33}$. The three columns in Table 11 correspond to these 
groups. The first column in Table 11 contains detailed information about   
the flavour structure of the RPV couplings in the 92 solutions with the 
hierarchy  $\lambda ^{\prime}_{133} >  \lambda ^{\prime}_{233},  
\lambda ^{\prime}_{333}$. The next few rows display the number of 
solutions with  predicted PBRs in different channels
  above the observable limits as given in Table 8. For example, 
the third row indicates that signals in  $e e + 2 jets$ and $e \tau +
2jets$ channels are observable with 10 fb $^{-1}$ of data in 56 
solutions. 
These channels, if observed, would further reveal that 
 $\lambda ^{\prime}_{133} >  \lambda ^{\prime}_{333} > \lambda 
^{\prime}_{233}$. On the other hand observable signals in $e e + 2 jets$ 
and $e \mu + 2 jets$ channels as 
given in the next row would indicate the  hierarchy  $\lambda 
^{\prime}_{133} 
>  \lambda ^{\prime}_{233} > \lambda^{\prime}_{233}$.  

If only one channel, say the $e e+2jets$, is observed one can conclude  
that  $\lambda ^{\prime}_{133} >>  \lambda ^{\prime}_{233}, \lambda
^{\prime}_{333}$ (see row 5). On the other the observation of the 
$e \mu + 2 jets$ signal only (see row 6) would indicate  $\lambda 
^{\prime}_{133} \approx  \lambda ^{\prime}_{233} >> \lambda
^{\prime}_{233}$. The channel $e \mu + 2 jets$ dominates over the
$e e + 2 jets$ or the $\mu \mu + 2 jets$ channel because of the factor 
of two which enhances the number of events when leptons of two 
different flavours with  all possible charge combinations are observed.
Finally the seventh row with `**' indicates that no signal can be 
observed with 
$\lum$ = 10 fb $^{-1}$

The information in the next two columns are presented following the
format
and similar inferences about the hierarchy of the  $\lambda 
^{\prime}_{i33}$ can be drawn from the lepton flavour content of
the final states. We have verified that for $\lum = 100 \ifb$ all
solutions well predict atleast one $P_{ij}$ above the corresponding
$P_{ij}^{min}$.

\section{Conclusion}

In conclusion we reiterate that the OSDL signals with same or
different flavours of leptons (e, $\mu$ or tau-jets) plus two additional jets 
arising from RPV decays of $\lstop$ - $\lstop^*$ pairs produced 
at the LHC would be a promising channel for probing the RPV 
coupling$\lambda^{\prime}_{i33}$ (see Eq. 1 and the discussions following it).
This is true in general if $\lstop$ 
happens to 
be the NLSP, which is a theoretically well motivated scenario.  
This signal is especially interesting in the context of RPV models
of neutrino mass. A part of our analysis (section 2), however, is fairly model independent since
the size of the signal is necessarily  
controlled by the production cross section of the $\lstop$ - $\lstop^*$ pair 
as given by QCD and the product branching ratio $P_{ij}$ (see Eq. 2).
The model independent
estimates of $P_{ij}^{min}$ (see Eq. 3) corresponding to 
observable signals for different $\mlstop$s (see Eq. 3) for an integrated 
luminosity of 10 $fb^{-1}$ are
presented in Table 7 and Table 8 using the Monte Carlo event generator Pythia. We have 
optimized the cuts for $\mlstop=240$ GeV. However, for even larger values of $\mlstop$
the signal efficiencies increase for the same set of cuts keeping the background events 
almost negligible. Top squark masses in the vicinity of 500 $\gev$ yield observable 
signals in this channel for realistic  models of $m_{\nu}$.
Although our calculations 
are based on LO top squark pair production cross sections we emphasise that the inclusion 
of NLO corrections are likely to yield even larger estimates of $P_{ij}^{min}$ 
as argued in section 2.  

We have further noted that inspite of the combinatorial backgrounds, the
invariant mass distribution of the lepton (e or $\mu$)-jet pair shows a peak at
$\mlstop$ (see Fig.1). This peak, if discovered, will clearly establish the
lepton number violating nature of the underlying interaction. This may not be possible if neutralino decays happen to be the only RPV signal.   

In models of $\nu $-mass, the underlying $\lambda^{\prime}$ couplings  
turn out to be very small. If $\lambda^{\prime}_{i33}$ contributes to the 
one loop $\nu$-mass matrix, it is typically of the order of 10$^{-4}$ - 
10$^{-5}$ due to constraints imposed by the $\nu$-oscillation data. Even if 
$\lambda^{\prime}$ is so small the RPV decay of the $\lstop$-NLSP may have sizable BRs
over a large region of the parameter space because the 
competing loop induced decay
(Eq. 4c) or the four body decay (Eq. 4d) of $\lstop$ also have suppressed widths. 
For the purpose
of illustration we have considered a specific model of $\nu$-mass\cite{abada}
with parameters constrained by the $\nu$-oscillation data. It is interesting to note that in 
this model most of the theoretically predicted $P_{ij}$'s (Eq. 2) for several representative 
choices of RPC parameters turn out to be larger than the  $P_{ij}^{min}$'s
 estimated in section 2 for $\lum$=
 10 $fb^{-1}$. For larger $\lum$ almost all solutions yield $P_{ij}$'s at 
 the observable level. 
 The relative size of the observed final states with various lepton 
 flavour contents will indicate the hierarchy among the 
 $\lambda^{\prime}_{i33}$s
 for different $i$'s.

{\bf Acknowledgement}:
AD and SP acknowledge financial support from  Department of Science and
Technology, Government of India under the project  No (SR/S2/HEP-18/2003).



\end{document}
